\documentclass[useAMS,usenatbib]{mn2e}
\newcommand{\dinbas}{DynBaS}
\newcommand{\dinbasuno}{DynBaS1D}
\newcommand{\dinbasdos}{DynBaS2D}
\newcommand{\dinbastres}{DynBaS3D}
\newcommand{\dinbasnd}{DynBaSND}

\newcommand{\starlight}{{\sevensize STARLIGHT}}
\newcommand{\ch} {$\chi^2_\nu$}

\usepackage{graphicx}
\usepackage{natbib}
\usepackage{amssymb}
\usepackage{lscape}
\usepackage{longtable}
\usepackage{color}
\title[An SSP Young Massive Cluster in NGC 34]{Constraining globular cluster formation through studies of young massive clusters - II. A Single Stellar Population Young Massive Cluster in NGC 34}
\author[I. Cabrera-Ziri et al.]{I. Cabrera-Ziri$^{1}$\thanks{ICZ: I.CabreraZiriCastro@2013.ljmu.ac.uk}, N. Bastian$^{1}$, B. Davies$^{1}$, G. Magris$^{2}$, G. Bruzual$^{3}$, F. Schweizer$^{4}$\\
$^{1}$ Astrophysics Research Institute, Liverpool John Moores University, 146 Brownlow Hill, Liverpool L3 5RF, UK\\
$^{2}$ Centro de Investigaciones de Astronom\'ia, AP 264, M\'erida 5101-A, Venezuela\\
$^{3}$ Centro de Radioastronom\'ia y Astrof\'isica, CRyA, UNAM, Campus Morelia, A.P. 3-72, C.P.58089 Michoac\'an, Mexico\\
$^{4}$ Carnegie Observatories, 813 Santa Barbara Street, Pasadena, CA 91101, USA\\
}
\begin{document}

\date{Accepted XXX. Received XXX; in original form XXX}

\pagerange{\pageref{firstpage}--\pageref{lastpage}} \pubyear{2014}

\maketitle

\label{firstpage}

\begin{abstract}

Currently there are two competing scenarios to explain the origin of the stellar population in globular clusters (GCs). The main difference between them is whether or not multiple events of star formation took place within GCs. In this paper we present the star formation history (SFH) of Cluster 1, a massive young cluster in NGC 34 $(\sim10^7\mbox{ M}_\odot)$. We use \dinbas, a spectrum fitting algorithm, to retrieve the SFH and find that Cluster 1 is consistent with a single stellar population of solar metallicity with an age of $100\pm30$ Myr and a mass of $1.9\pm0.4\times10^7\mbox{ M}_\odot$. These results are in conflict with the expectations/predictions of the scenarios that invoke extended or multiple episodes within  30--100 Myr of the initial star-formation burst in young massive clusters.



\end{abstract}


\section{Introduction}
\label{sec:intro}

The classical notion of globular clusters (GCs) being simple stellar populations (SSPs) has been challenged by the presence of chemical anomalies and multiple sequences in the colour--magnitude diagrams (CMD) of GCs. The chemical anomalies are present only in light elements (namely C, N, O, Na and Al - e.g. \citealt{Carretta:2009p2165}) and are generally found only within GCs and not in the field population (e.g. \citealt{Martell:2011p2166}). To date, only a handful of clusters have been found with significant Fe spreads among their stellar populations (e.g. $\omega$ Centauri, Terzan 5, M52, M22 and NGC1851 - e.g. \citealt{2014arXiv1401.4323M} and references therein). Additionally, significant spreads in He abundance within GCs  have been proposed to explain multiple main sequences and turn-offs, as well as the shape of the horizontal branch, in colour--magnitude diagrams of some GCs (e.g. \citealt{Milone:2012p2160}).

Most models that attempt to explain the chemical anomalies and CMD morphology observed in GCs assume that these features are the product of multiple generations of stars.  The basic idea is that a second generation of stars is created from the chemically processed ejecta of some very precise kinds of stars from the first generation (\emph{polluter} stars). Stars that have been suggested to be \emph{polluters} include: Asymptotic Giant Branch (AGB) stars (e.g. \citealt{Dercole:2008p2154}), fast rotating massive stars (also known as spin-stars e.g. \citealt{Decressin:2009p2155}), and massive stars in interacting binary systems \citep{DeMink:2009p2156}.

All these multiple-population scenarios do well reproducing many of the observed anomalies mentioned before, and they predict that star clusters forming today should undergo a second generation of star formation. If spin-stars or massive interacting binaries are the source of the enriched material, then the second generation is expected to form within $\sim10$~Myr of the first generation.  Alternatively, if AGB stars are the source, a difference of $30-200$~Myr between the 1st and 2nd generation is expected (e.g. \citealt{Conroy:2011p1997}). Recently, an alternative scenario has been proposed that does not invoke multiple star formation events within massive clusters. In this scenario, \cite{Bastian:2013p2152} suggest that the chemically enriched material is ejected by spin stars or high mass interacting binaries, and is accreted onto circumstellar disks of pre-main-sequence low mass stars of the same generation.

Additionally it has been suggested that the observed extended main sequence turn-offs (eMSTO) and ``dual red clumps'' observed in intermediate age (1--2 Gyr) Small and Large Magellanic Cloud (LMC) clusters may be the product of extended (200--500 Myr) star formation events (e.g. \citealt{Mackey2007,Goudfrooij:2009p2234,Goudfrooij:2011p2248,Goudfrooij:2011p2254,Milone:2009p2291,Rubele:2013p2304}). Some studies propose a common evolution of these intermediate age clusters with GCs (e.g. \citealt{Conroy:2011p1997}). On the other hand, there are some claims for the opposite, for example \cite{Mucciarelli:2008p2339} indicated that the eMSTO of intermediate-age clusters were not related to the multiple stellar populations seen in globular clusters, due to the lack of abundance spreads between the stars of the younger clusters. Alternatively, different mechanisms have been put forward to explain such anomalies in intermediate age clusters e.g. stellar rotation (e.g. \citealt{Bastian:2009p2488,Yang:2013p2467}) or interacting binaries (e.g. \citealt{Yang2011}).

\cite{Bastian:2013p2152} argue that if such extended (or multiple) star formation events took place in these intermediate age clusters and GCs, it would be expected that younger ($<500$ Myr) massive clusters should be currently forming stars. To test this, \cite{Bastian:2013p2199} studied the CMD of two young (180--280 Myr) massive ($\sim10^5$ M$_\odot$) clusters in the LMC and assessed an upper limit of 35 Myr for the possible age spread in these clusters. Also \cite{Bastian:2013p2022} presented a catalog containing more than 100 young (10--1000 Myr) massive ($10^4$--$10^8$ M$_\odot$) clusters where they do not find evidence of any ongoing star formation within the clusters, and concluded that any extended star formation within clusters lasting for hundreds of Myr are ruled out at high significance (unless strong stellar initial mass function -IMF- variations are invoked).  Their study was sensitive to $\sim2$\% of the current cluster mass being formed today.  If such extended ($200-500$~Myr) star formation events were common, the authors estimate that roughly 50\% of their sample should have shown evidence for ongoing star-formation.

In this work, we approach the problems of the origin of multiple populations in GCs and eMSTO/``dual red clumps'' detected in intermediate age clusters by analysing the integrated spectrum of a young massive star cluster, looking for evidence for multiple events of star formation within this cluster. The cluster we chose for this initial study is young ($\sim$150~Myr), is found in the wet-merger galaxy NGC 34 and does not show any evidence for ongoing star-formation, based on the lack of optical emission lines in its spectrum (e.g., \citealt{Schweizer:2007p2018}; \citealt{Bastian:2013p2022}). This young globular cluster (Cluster 1, hereafter) has an estimated mass of about 15--20$\times10^6\mbox{ M}_\odot$ \citep{Schweizer:2007p2018}, which is 3--4 times more massive than that of $\omega$ Centauri, the most massive GC in the Galaxy. The fact that Cluster 1 is so massive and young makes it rather suitable to probe both families of formation scenarios, given that it can easily retain the ejecta of the polluter stars of the first generation, and we should be able to find evidence of a second generation of stars if a secondary burst has already taken place in the cluster.

The paper is organised as follows:  In \S \ref{sec:data} we present the optical spectrum of Cluster 1 and in \S \ref{sec:dinbas} we introduce the fitting method and models used in the SFH analysis.  The degeneracies and uncertainties in the fits are discussed in \S \ref{sec:uncertainties}, and we discuss our results and present our conclusions in \S \ref{sec:discussion} and \S\ref{sec:conclusions}, respectively.

\section{Data}
\label{sec:data}


We analyse the integrated spectrum of one of the most massive clusters in NGC 34 (Mrk 938), Cluster 1. The spectrum was obtained by \citealt{Schweizer:2007p2018} (hereafter, SS07) with the Low Dispersion Survey Spectrograph (LDSS-2) of the Baade 6.5 m telescope at Las Campanas. This spectrum has a spectral resolution of $\sim$ 5.3 {\AA} at 5000 {\AA}, and a wavelength coverage of about 3700--6850 {\AA} (see SS07 for more details regarding instrument settings and reduction).

We note that the spectrograph used (LDSS-2) does not have an Atmospheric Dispersion Corrector and  the targets were observed with a \textit{non-parallactic slit} at airmasses of ca 1.1--1.2 (in order to include 2 clusters at a time for each slit placing).  Therefore, differential refraction might have led to wavelength-dependent light losses and the wrong continuum shape. 

Since the continuum of the spectrum may not be representative of the actual flux levels, the only SFH diagnostics of the spectrum that we can use are the line-to-continuum ratios of absorption features. Therefore we normalised the continuum for our SED fitting. To obtain the continuum we ran a median filter of 100 {\AA} width over the cluster spectrum, masking 2,000 km/s around the core of each Balmer line. For the blue end of the spectrum (wavelengths on the blue side of H$\epsilon$) the pseudo-continuum was not properly reproduced with the median filter, and a handcrafted continuum was used over this wavelength range in order to improve the continuum fit. 

We tested that our results are not affected by the exact choice of the continuum normalization method (see also \S \ref{sec:uncertainties}).

\begin{figure*}
\includegraphics[width= 180mm,height=84mm]{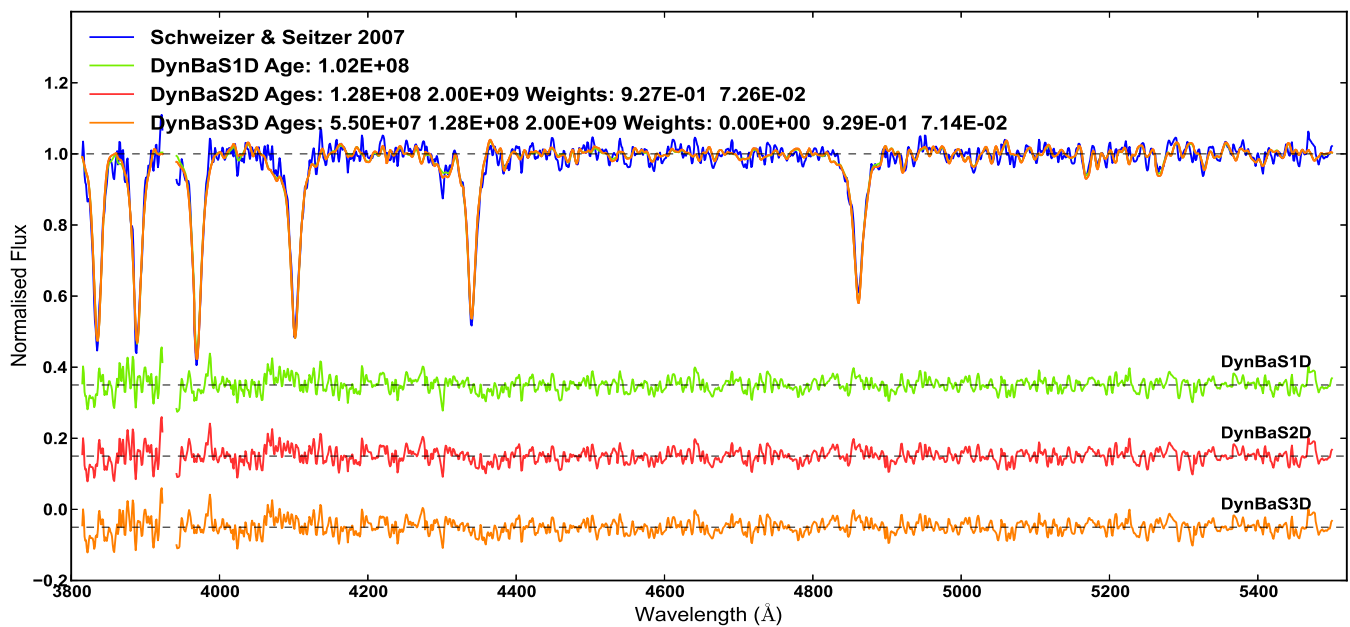}
\caption{\dinbasuno, \dinbasdos\ and \dinbastres\ fits to the continuum-normalised spectrum of Cluster 1. The \dinbas\ fits presented here were obtained using BC03 models of solar metallicity. On the bottom we plot the residuals (data - \dinbas\ fits) in the same vertical scale with 3 different offsets for clarity. The three solutions provide virtually the same SED. Note that the Ca{\sc ii} K (3933 \AA) line has been masked for the fit.}
\label{dinbasfit}
\end{figure*}

Once the (pseudo)continuum was found for the cluster SED, we divided the observed spectrum by the continuum to produce the normalised spectrum employed in our analysis and shown in Fig.~\ref{dinbasfit}.




%

\section{\dinbas\ fitting}
\label{sec:dinbas}

We make use of a Dynamical Basis Selection spectral fitting algorithm (\dinbas) originally
developed to recover the SFH of galaxies \citep{Magris2014}. Most SED fitting algorithms, e.g. MOPED, VESPA, STECMAP, \starlight, ULySS \citep{Heavens:2000p2501,Tojeiro:2007p9, Ocvirk:2006p2527, CidFernandes:2005p4, Koleva:2009p2546}, use a fixed spectral basis to fit all target spectra. Instead, in \dinbasnd\footnote{\dinbasnd\ denotes that a basis of $N$ spectra is used in the fits.} for each target spectrum a different and dynamically selected basis of $N$ model spectra is used to obtain an optimal fit to the target spectrum. In practice, we fit the target spectrum using all possible combinations of $N$ model spectra and store the resulting \ch\ for each solution. The \dinbasnd\ solution is then the one with the minimum \ch, subject to the condition that the weight $a_{i}$ assigned to the $ith$ spectrum in the basis obeys $a_{i} \ge 0$ for $i = 1,..., N$. As argued by \cite{Magris2014}, \dinbasnd\ for $N$ = 2 and 3 provide excellent fits to the target spectra, and the residuals of the recovered physical parameters for the target galaxies are less biased than for fixed-age, rigid basis methods.

These features make \dinbas\ a
suitable tool to analyse the integrated spectra of young clusters since their SFHs are expected to comprise just a few star formation bursts of short length (e.g. \citealt{Gratton:2012p2005}). This gains an edge for the study of these clusters over conventional fixed basis SED fitting codes, since the latter may introduce a great number of artificial components (i.e. ages) to the fit while exploring the vast parameter space that is the set of ages of current stellar population synthesis models. 
For example \cite{CidFernandes:2010p2159} and \cite{Dias:2010p2158} found unphysical solutions for the multiple-populations fits to young and old stellar clusters, e.g. age differences of the order of 10 Gyr within clusters. On the other hand, \dinbas\ will adapt to each target spectrum's peculiarities, fitting it with the best linear combination of $N$ components from the whole set of ages available in the models, reducing the number of artificial components and simplifying the analysis.

\begin{figure*}
\includegraphics[width= 58mm]{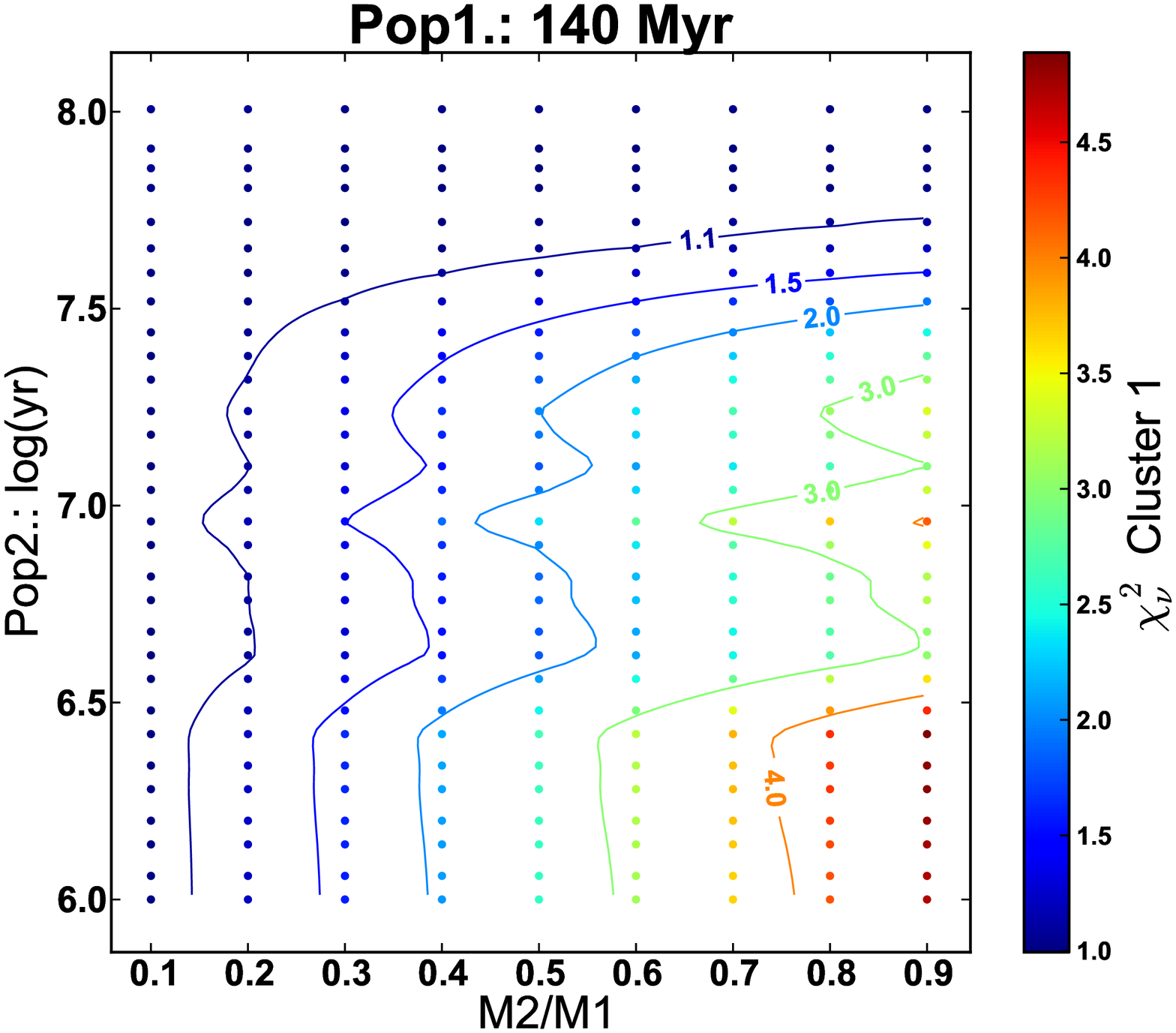}
\includegraphics[width= 58mm]{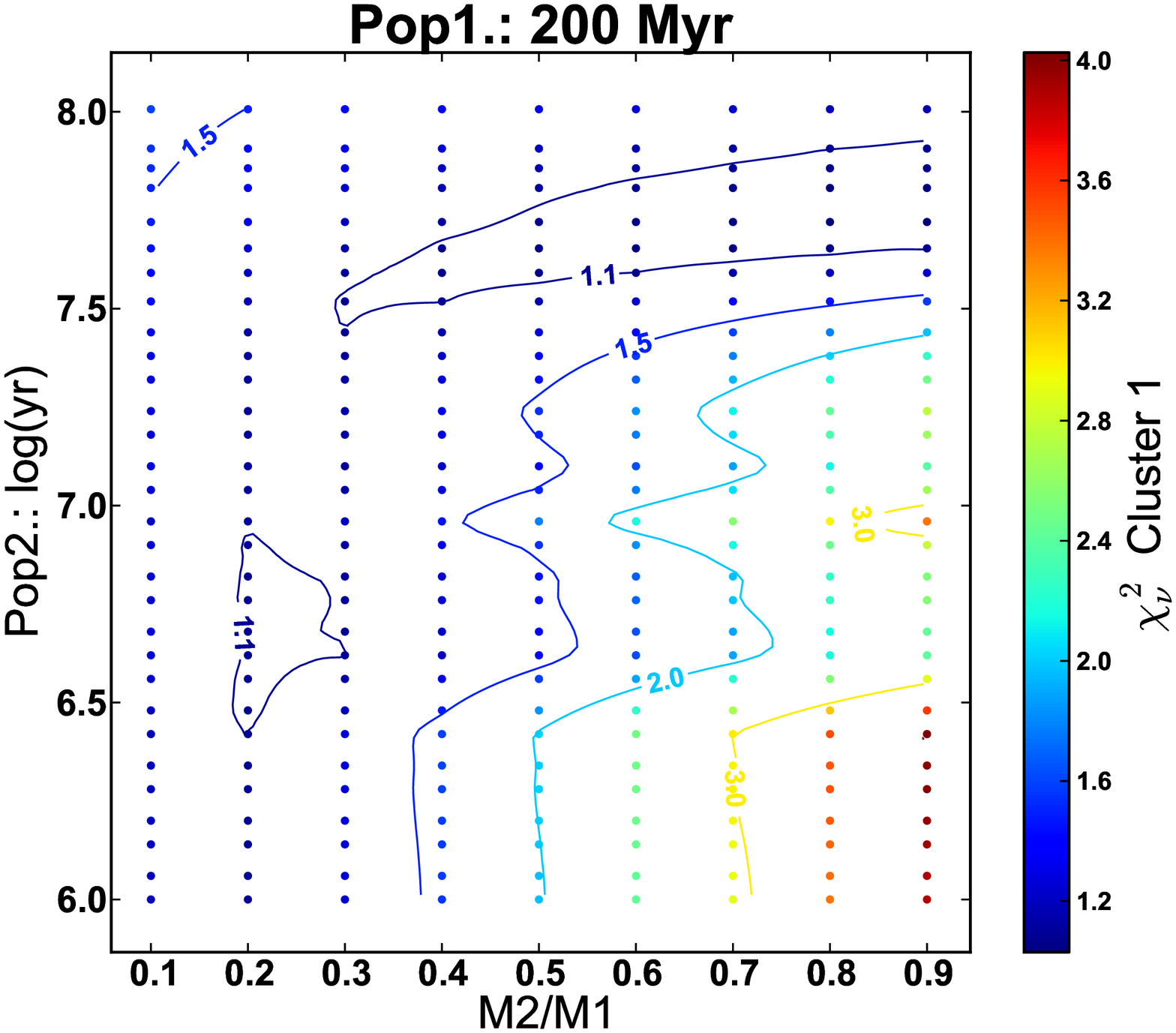}
\includegraphics[width= 58mm]{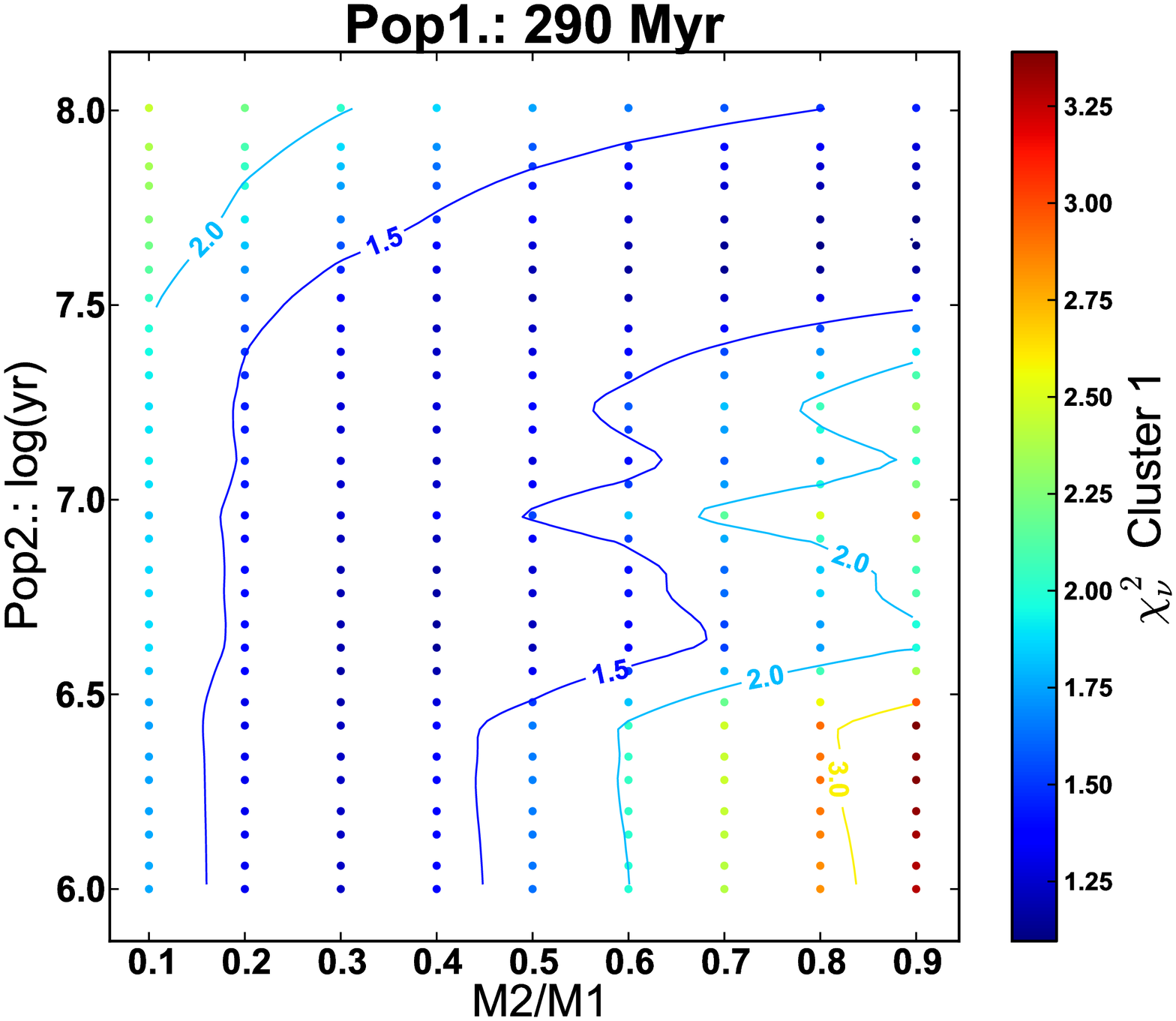}
\caption{
Results of fitting the normalised spectrum of Cluster 1 with each element of three representative grids of synthetic cluster spectra. The Pop. 1 ages are (from left to right) 140, 200, and 290 Myr. The vertical axis represents the age of the secondary (less massive) population, while the horizontal axis denotes the mass ratio between the first (older and more massive) and second population. Note that the $\chi_\nu^2$ range changes for every panel, hence the colour coding does not represent the same values in different panels. For $\chi_\nu^2>1.1$ we can spot the differences by eye between the synthetic clusters and the data, so those solutions are immediately excluded.}
\label{grids}
\end{figure*}
\subsection{Star Formation History}
\label{sec:sfh}


We used \citealt{bc03} (hereafter BC03) stellar population models for the SED fitting, assuming a \cite{Chabrier:2003p97} IMF, computed using ``Padova 1994" evolutionary tracks \citep{Alongi:1993p2051,Bressan:1993p2083,Fagotto:1994p2104,Fagotto:1994p2101,Girardi:1996p2140} and the stellar library STELIB \citep{LeBorgne:2003p2040}. We normalised the continuum for each SSP comprised in the models, using again a median filter with the same width and mask that we used for the observed spectrum, but on the blue end we fit a cubic spline between the pseudo-continuum between the Balmer lines. Finally, we divided each SED by its respective continuum.

For the fits we used the entire range of ages contained in the models, ranging from $10^5$ to $2\times10^{10}$ yr. The fits were performed between 3815--5500 {\AA}, comprising the region most sensitive to age in the optical spectra \citep{Schweizer:1998p2017}, including most Balmer lines and some important metal indices (e.g. Fe5015, Fe5270, Fe5335 and Mg$b$) for metallicity estimation. For the fits we masked the Ca{\sc ii} K line, since it might be contaminated by interstellar absorption within NGC 34 (SS07).

The cluster velocity dispersion was treated as a free parameter during the fits, as a means to emulate the true dispersion and as a correction factor to downgrade the resolution of the models to match the resolution of the data.

We estimated the S/N of the Cluster 1 spectrum to be 48 between 5000-5800 {\AA} and we assumed it to be constant over the entire spectrum. Since we did not have a proper error spectrum for the fits, we used the observed flux divided by the S/N of our data as error spectrum. Because the error spectrum is not real, the $\chi_\nu^2$ are not genuine either but they are comparable between different solutions that used the same error spectrum.

The $\chi_\nu^2$ obtained in our \dinbasuno\ fit was 1.27. For comparison reasons, we lowered the S/N estimated before dividing it by the value of our $\chi_\nu$ $(\sqrt{1.27})$, then we calculated a new error spectrum for this value. This resulted in a value of $\chi_\nu^2=1$ for our \dinbasuno\ solution using the new error spectrum, and made $\chi_\nu^2$ values from the grids in \S \ref{sec:uncertainties} easier to compare with \dinbas\ results.



Figure~\ref{dinbasfit} shows the DynBaS1D, DynBaS2D, and DynBaS3D fits to our Cluster 1 SED. \dinbasuno\ yields an age of 100 Myr. For \dinbasdos\ we find ages of 130 Myr (quite similar to \dinbasuno\ solution) and 2 Gyr, corresponding to 93\% and 7\% of the total mass of the cluster, respectively. Finally, the 3-component fit \dinbastres\ yields ages of 55 Myr, 130 Myr and 2 Gyr for stellar populations containing $<$0.01\%, 93\% and 7\% of the cluster mass, respectively.


As can be seen in the residuals plotted at the bottom of  Fig.~\ref{dinbasfit}, the differences between the three fits are very small, and in principle each is as valid as the others. However, we can rule out the multiple population solutions (\dinbasdos\ and \dinbastres) given the lack of physical meaning of the results.  For example, we note that  ``very old" ($>$1 Gyr) populations in these clusters are not expected to exist, and it would imply that a low-mass cluster existed for more than a Gyr before a second generation formed within the cluster, with $\sim13$ times the mass of the initial population.

In \S \ref{sec:uncertainties} we will place more constraints on  the multiple population solutions and estimate the uncertainties in our age determination.

\subsection{Metallicity and Mass}

In a previous work, SS07 estimated that Cluster 1 has an age of $150\pm20$ Myr and solar metallicity through the analysis of Lick line indices. Here we made various fits assuming metalicities of $Z=0.4$, 1 and 2.5$Z_\odot$ ($[Z]=-0.4$, 0 and 0.4) for the models, and we found that the best fits to spectra (specifically, the 5100--5400 {\AA} region which hosts a number of important metallicity indicators, including Mg$b$, Fe5270, Fe5335 - \citealt{Gonzalez:1993p2231}) were with the $Z_\odot$ templates (values reported in \S \ref{sec:sfh}). Given this result, we restrict ourselves to the $Z_\odot$ models for the rest of the analysis.

%

%


The mass was estimated in the standard way, through a comparison between the observed cluster luminosity (corrected for distance and extinction) and predictions of SSP models for the corresponding age (which assume a metallicity and stellar initial mass function).  
We adopt the photometry of SS07 (V=19.38), an extinction of $A_V = 0.1$, and a distance of 85.2~Mpc, to derive an absolute V-band magnitude of $M_V=-15.36$ for this cluster.  Comparing this to predictions from the BC03 models for solar metallicity, and an age of 100 Myr (adopting a \citealt{Chabrier:2003p97} IMF) we estimate the mass of Cluster 1 to be $1.9\times10^7\mbox{ M}_\odot$.  An uncertainty of 10\% on the distance leads to an uncertainty of 20\% on the estimated mass.  Uncertainties related to adopting specific SSP models are also at the $20-30$\% level.

\section{Degeneracies and Uncertainties}
\label{sec:uncertainties}

To assess possible degeneracies in our results (i.e. if other combinations of multiple
populations reproduce the Cluster 1 SED equally well as our best solution) we performed theoretical experiments over grids of synthetic multiple-population clusters.


The grids were made up of synthetic cluster spectra for two events of star formation. These spectra were built using the same BC03 models we used for our \dinbas\ fits. Each grid consisted of synthetic clusters with the same older population (Pop. 1 from here on) and different younger populations (Pop. 2). The masses of Pop. 2 could take values ranging from 10\% to 90\% of the mass of Pop. 1, and for the second population we allowed ages between 1 and 100 Myr distributed almost uniformly in log space. For Pop. 1 we used a wide extent of ages with very small time steps between them, creating a finely sampled parameter space (namely 140, 200, 290, 400, 510, 570, 720, 810 and 900 Myr).

We then applied to each synthetic cluster SED a Gaussian filter with the same velocity dispersion that \dinbas\ had recovered in the $Z_\odot$ fit. Then for each element (synthetic 2-population cluster) of every grid we fitted the spectrum of Cluster 1 by minimising the $\chi_\nu^2$ between them using a standard least-squares algorithm. The values of $\chi_\nu^2$ were computed using the same error spectrum we used to normalise the $\chi_\nu^2$ in the previous section.

%

Figure~\ref{grids} presents the results of fitting Cluster 1 to three representative grids. In the figure we colour coded the solutions as a function of their $\chi_\nu^2$. We found that for fits with $\chi_\nu^2>1.1$ it is possible to distinguish by eye that the spectral fits are poor (i.e. fail to reproduce the depths/profiles of the Balmer lines). The contours denote constant values of $\chi_\nu^2$. Note how the areas with small values of $\chi_\nu^2$ rapidly shrink when we increase the age of Pop. 1 (areas enclosed by dark blue contours get smaller with older massive population). For grids with Pop. 1 values older than 200 Myr we can easily rule out the solutions at high significance.



We know that solutions with Pop. 2 ages younger than 10 Myr are not possible on account of the lack of emission lines produced by ionised gas.  Such line emission can be detected down to limits of $\sim$ 2\% of the Pop. 1 mass \citep{Bastian:2013p2022}. Hence, we can reject all solutions with these ages. As Fig.~\ref{grids} shows, the ``good fits'' are only for massive and older Pop. 2 ages (Pop. 2 $\simeq$ Pop 1.) or very young Pop. 2 and with small mass ratios (M$_2\ll$ M$_1$) when we increase the age of the main population Pop. 1. 

\begin{figure}
\includegraphics[width= 84mm]{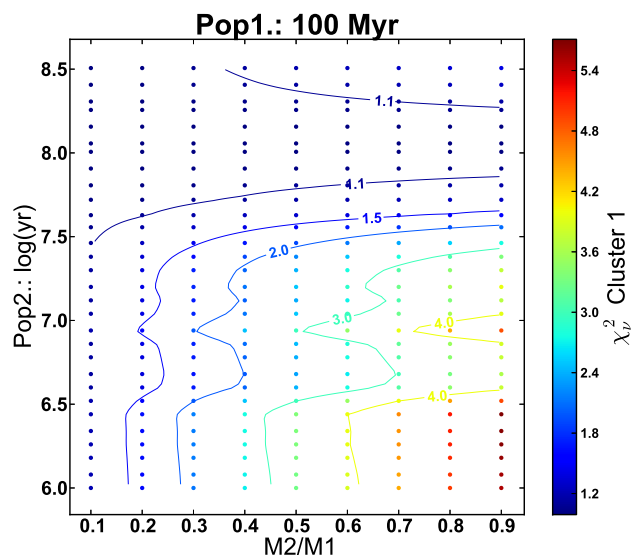}
\caption{
Results of fitting the normalised spectrum of Cluster 1 with each element of the 100 Myr extended grid. This grid show that the solutions are consistent with the \dinbasuno\ results (i.e. an SSP of $\sim100$ Myr).}
\label{grids100}
\end{figure}

We also carried out an additional experiment with a grid containing a Pop. 1 of 100 Myr (massive), but for this grid we allowed the Pop. 2 (less massive) ages to reach $\sim300$ Myr (i.e. to exceed the age of Pop. 1), with the results shown in Fig.~\ref{grids100}. In this figure we can see a tendency where the best solution (fits) is the 100 Myr SSP (row with Pop. 2 $\log{\mbox{(yr)}} = 8.0$) with the quality of the fits (hence the likelihood of the solutions) gradually degrading in the direction of old massive Pop. 2, or young and less massive ones.  With a main population of 100~Myr, the spectrum does not change significantly when adding a small amount of mass in a secondary population  with older/younger ages, or a large amount of mass with an age close to 100~Myr. From Figs.~\ref{grids} \&\ \ref{grids100} we can see that there are regions of parameter space where a second generation could be hidden.  By looking at other young massive clusters with ages between 12 and 500~Myr, we should be able to sample all of these regions, and remove any degeneracies.  This will be carried out in a future work (Cabrera-Ziri et al.~in prep.).



We emphasize that while experimenting with different ways to define the continuum in our data and models, defining different wavelength ranges for the fits, and assuming different S/N for the data \emph{we always found that one-population solutions performed better than multiple-population solutions}. Almost all of these solutions lay within 10\% of the reported age (100 Myr). We found the same behaviour (SSP solutions) when we performed the same trials with different metallicities.

\section{Discussion}
\label{sec:discussion}

The tests performed (shown in Figs.~\ref{grids} \& \ref{grids100}) show that it is possible to construct a spectrum similar to that of Cluster 1 by combing SSP models of specific ages and mass fractions.  However, we note that the \dinbasuno\ solution to the observed spectra is preferred to the multiple generation model spectra. We use Fig. \ref{grids100} to estimate an error for  \dinbasuno\ solution of 30 Myr i.e. an age for Cluster 1 of $100\pm30$ Myr and a mass of $1.9\pm0.4\times10^7\mbox{ M}_\odot$. As an exploratory theoretical experiment, we also performed \dinbas\ fitting for the grids, and we found that multiple-population solutions (i.e. \dinbasdos\ and \dinbastres) did better than \dinbasuno\ solutions. From this we conclude that if multiple populations like the ones of our grids were actually present in this cluster, we would have been able to recover them with \dinbas. In a future work, we will present these tests and a detailed analysis of them.  Additionally, we will assess the uncertainties introduced with the choice of selected stellar population models.

\section{Conclusions}
\label{sec:conclusions}

By fitting the normalised spectrum of Cluster 1 in NGC 34 with model SSP spectra, we have determined an age of $100\pm30$ Myr for the cluster and estimated a mass of  $1.9\pm0.4\times10^7\mbox{ M}_\odot$, based on published photometry and SSP models for this age. We do not find evidence for multiple star formation episodes, and we can confidently rule out the presence of a 2nd generation of stars for ages outside the range from 70 to 130 Myr with mass ratios between the second and first generation greater than 0.1. These results are consistent with GC formation scenarios where multiple generations of stars are separated by $<30$ Myr in age (e.g. \citealt{Decressin:2009p2155,DeMink:2009p2156}) or scenarios that do not invoke multiple star forming events (\citealt{Bastian:2013p2152}).



Our results do not support any GC formation scenarios that involve multiple generations of stars separated by $>30$ Myr in age. However, it is still possible that a secondary burst might happen in the future (i.e. with an age difference between the first and second generation of stars that is greater than 100 Myr). To improve our understanding of how GCs form, further spectroscopic studies of young massive clusters covering a wide range of ages are needed. In a separate paper, Cabrera-Ziri et al. (in prep.) determine the SFH of 6 young (12--500 Myr) massive ($>10^6\mbox{ M}_\odot$) clusters from an ongoing spectroscopic survey.



These result are consistent with the findings of \cite{Bastian:2013p2199}, who do not find any large age spreads in young massive LMC clusters, and they also disagree with GC formation scenarios that predict extended SFHs (e.g. \citealt{Conroy:2011p1997}).

Finally, we conclude that \dinbas\ capabilities (i.e. SED fitting of just a few ages) are ideal for the study of the integrated spectra of young clusters, given that they reduce significantly the amounts of non-genuine components (i.e. ages) compared to traditional SED fitting algorithms, consequently simplifying the analysis of the results.

\section{Acknowledgements}

GB acknowledges support from the National Autonomous University of M\'exico, through grant IB102212-RR182212. NB is partially funded by a Royal Society University Research Fellowship.  We thank the Aspen Center for Physics and the NSF Grant \#1066293 for hospitality during the conception of this project.

\bibliographystyle{apj}
\bibliography{latex_references}

\end{document}